\documentstyle[aas2pp4]{article}

\begin{document}

%-------------------- mathematics ---------------------------------
\def\la{\mathrel{\mathchoice {\vcenter{\offinterlineskip\halign{\hfil
$\displaystyle##$\hfil\cr<\cr\sim\cr}}}
{\vcenter{\offinterlineskip\halign{\hfil$\textstyle##$\hfil\cr
<\cr\sim\cr}}}
{\vcenter{\offinterlineskip\halign{\hfil$\scriptstyle##$\hfil\cr
<\cr\sim\cr}}}
{\vcenter{\offinterlineskip\halign{\hfil$\scriptscriptstyle##$\hfil\cr
<\cr\sim\cr}}}}}
\def\ga{\mathrel{\mathchoice {\vcenter{\offinterlineskip\halign{\hfil
$\displaystyle##$\hfil\cr>\cr\sim\cr}}}
{\vcenter{\offinterlineskip\halign{\hfil$\textstyle##$\hfil\cr
>\cr\sim\cr}}}
{\vcenter{\offinterlineskip\halign{\hfil$\scriptstyle##$\hfil\cr
>\cr\sim\cr}}}
{\vcenter{\offinterlineskip\halign{\hfil$\scriptscriptstyle##$\hfil\cr
>\cr\sim\cr}}}}}
\let\lsim=\la
\let\gsim=\ga

%------------------------------------------------------------------

\title{
Statistical Properties of Bright Galaxies in the SDSS Photometric 
System
\footnote{Based on observations obtained 
with the Sloan Digital Sky Survey}
}

\author{
Kazuhiro Shimasaku\altaffilmark{\ref{UTokyo}},
Masataka Fukugita\altaffilmark{\ref{CosmicRay},\ref{IAS}},
Mamoru Doi\altaffilmark{\ref{UTokyoMitaka}},
Masaru Hamabe\altaffilmark{\ref{UTokyoMitaka}},
Takashi Ichikawa\altaffilmark{\ref{Tohoku}},
Sadanori Okamura\altaffilmark{\ref{UTokyo}},
Maki Sekiguchi\altaffilmark{\ref{CosmicRay}},
Naoki Yasuda\altaffilmark{\ref{NAOJapan}},
Jon Brinkmann\altaffilmark{\ref{APO}},
Istv\'an Csabai\altaffilmark{\ref{JHU},\ref{Eotvos}},
Shin-Ichi Ichikawa\altaffilmark{\ref{NAOJapan}},
$\check {\rm Z}$eljko Ivezi\'c\altaffilmark{\ref{Princeton}},
Peter Z. Kunszt\altaffilmark{\ref{JHU}},
Donald P. Schneider\altaffilmark{\ref{PennState}},
Gyula P. Szokoly\altaffilmark{\ref{Potsdam}},
Masaru Watanabe\altaffilmark{\ref{ISAS}},
Donald G. York\altaffilmark{\ref{Chicago}},
}

\newcounter{address}
\setcounter{address}{1}
\addtocounter{address}{1}
\altaffiltext{\theaddress}{Department of Astronomy 
and Research Center for the Early Universe, School of Science, 
University of Tokyo, Tokyo, 113-0033 Japan
\label{UTokyo}}
\addtocounter{address}{1}
\altaffiltext{\theaddress}{Institute for Cosmic Ray Research, 
University of Tokyo, Kashiwa 277-8582, Japan
\label{CosmicRay}}
\addtocounter{address}{1}
\altaffiltext{\theaddress}{Institute for Advanced Study, Olden Lane,
Princeton, NJ 08540
\label{IAS}}
\addtocounter{address}{1}
\altaffiltext{\theaddress}{Institute of Astronomy, 
School of Science, University of Tokyo, Mitaka, Tokyo 181-0015, Japan
\label{UTokyoMitaka}}
\addtocounter{address}{1}
\altaffiltext{\theaddress}{Astronomical Institute,
Tohoku University, Sendai 980-8578 Japan
\label{Tohoku}}
\addtocounter{address}{1}
\altaffiltext{\theaddress}{National Astronomical Observatory, 
2-21-1, Mitaka, Tokyo 181-8588, Japan
\label{NAOJapan}}
\addtocounter{address}{1}
\altaffiltext{\theaddress}{Apache Point Observatory, P.O. Box 59,
Sunspot, NM 88349-0059
\label{APO}}
\addtocounter{address}{1}
\altaffiltext{\theaddress}{
Department of Physics and Astronomy, The Johns Hopkins University,
   3701 San Martin Drive, Baltimore, MD 21218, USA
\label{JHU}}
\addtocounter{address}{1}
\altaffiltext{\theaddress}{Department of Physics of Complex Systems,
E\"otv\"os University,
   P\'azm\'any P\'eter s\'et\'any 1/A, Budapest, H-1117, Hungary
\label{Eotvos}}
\addtocounter{address}{1}
\altaffiltext{\theaddress}{Princeton University Observatory, 
Princeton, NJ 08544
\label{Princeton}}
\addtocounter{address}{1}
\altaffiltext{\theaddress}{Department of Astronomy and Astrophysics,
The Pennsylvania State University,
University Park, PA 16802
\label{PennState}}
\addtocounter{address}{1}
\altaffiltext{\theaddress}{Astrophysikalisches Institut Potsdam, 
An der Sternwarte 16, 14482 Potsdam, Germany
\label{Potsdam}}
\addtocounter{address}{1}
\altaffiltext{\theaddress}{The Institute of Space and Astronautical 
Science, Sagamihara, Kanagawa 229-8510, Japan
\label{ISAS}}
\addtocounter{address}{1}
\altaffiltext{\theaddress}{University of Chicago, 
Astronomy \& Astrophysics Center, 
5640 S. Ellis Ave., Chicago, IL 60637
\label{Chicago}}

%------------------------------------------------------------------

\begin{abstract}

We investigate the photometric properties of 456 bright galaxies 
using imaging data recorded during the commissioning phase of 
the Sloan Digital Sky Survey (SDSS). 
Morphological classification is carried out by correlating
results of several human classifiers.
Our purpose is to examine the statistical properties of color indices,
scale lengths, and concentration indices
as functions of morphology for the SDSS photometric system.
We find that $u'-g'$, $g'-r'$, and $r'-i'$ colors of SDSS galaxies
match well with those
expected from the synthetic calculation of spectroscopic energy
distribution of template galaxies and with those transformed
from $UBVR_CI_C$ color data of nearby galaxies.
The agreement is somewhat poor, however, for $i'-z'$ color band 
with a discrepancy of $0.1-0.2$ mag.
With the aid of the relation between surface brightness and radius 
obtained by Kent (1985), 
we estimate the averages of the effective radius of early type
galaxies and the scale length of exponential disks both to be 2.6 kpc
for $L^*$ galaxies.
We find that the half light radius of galaxies depends slightly
on the color bands, consistent with the expected distribution of
star-forming regions for late type galaxies and with the known
color gradient for early type galaxies.
We also show that the (inverse) concentration index, defined by
the ratio of the half light Petrosian radius to the $90\%$
light Petrosian radius, 
correlates tightly with the morphological type;
this index allows us to classify galaxies into early (E/S0) and
late (spiral and irregular) types, allowing for a 15-20\% contamination
from the opposite class compared with eye-classified morphology.

\end{abstract}

\keywords{galaxies:fundamental parameters---galaxies:photometry}

%------------------------------------------------------------------

\section{Introduction}

The Sloan Digital Sky Survey (SDSS; York et al 2000) 
has started producing both photometric and spectroscopic data. 
It is expected that the SDSS will produce
a very large and homogeneous database for research 
on the nature of galaxies.
Photometric observations are made with the new five color 
bands ($u'$, $g'$, $r'$, $i'$, and $z'$) 
that divide the entire range from 
the atmospheric ultraviolet cutoff to the sensitivity limit 
of silicon CCD into five, essentially non-overlapping,
passbands (Fukugita et al. 1996). 
These passbands are chosen on the basis of their astrophysical
merits, but this specific choice makes it difficult to compare
or combine the SDSS data with those obtained with more
conventional photometric systems. 

In this paper, we present results from a brief analysis 
for properties of galaxies using a small set of data for bright 
galaxies in the new photometric system. 
The present results 
have already been used to construct empirical galaxy models for
simulations that have extensively been used to tune 
the SDSS project software (Lupton et al. unpublished) and to aid 
galaxy science (Yasuda et al. 2000).

To study properties of galaxies, it is essential to classify the
galaxies into morphological types, since different morphological types
exhibit distinctly different astrophysical properties, 
reflecting the different histories of the formation and
evolution of galaxies.
A number of attempts have been proposed 
for automated classification of morphologies 
(e.g., Doi, Fukugita, \& Okamura 1993; 
Lahav et al. 1996; Abraham et al. 1994, 1996), 
but visual classification still serves as the most reliable method 
when we adopt the Hubble classification (Sandage 1961) for 
galaxies with large apparent sizes.
This visual inspection procedure, 
which is of course very labor-intensive, limits the size of our sample. 
While we are conducting an attempt to produce a large-scale
eye-classified sample, we present here 
an analysis we have carried out as our initial study.

In this paper, we focus our considerations on three statistical
quantities: color, effective size, and concentration parameter. 
Galaxy colors are an important quantity that characterizes stellar
contents of galaxies.
The colors at zero redshift are also
used as a fiducial zero point to study statistical properties of 
faint galaxies and their evolution.
An important test is to confirm whether the 
colors obtained with SDSS photometry are 
consistent with those expected from broad band photometry using the 
conventional passbands, $B$, $V$, $R$, and $I$, 
and with those calculated from 
template spectroscopic energy distributions of galaxies.
This is an initial step in establishing 
a photometric system where the spectrophotometrically synthesized flux 
matches closely with the broad band flux, which is
one of the goals of SDSS photometry. 

The second statistical quantity is the distribution 
of effective sizes of galaxies. The de Vaucouleurs profile 
(de Vaucouleurs 1948) is
usually characterized by the half light radius $r_e$ 
and the exponential disk by the scale length $h$. 
These parameters set the fundamental length
scales of galaxies.
How these quantities scale with magnitude is
indispensable information for construction of simulations of galaxies, 
in particular the study of the detectability of 
galaxies at faint magnitudes and the
performance of star galaxy classification algorithms, 
both of which depend on the apparent
sizes of galaxies. It is also interesting to ask whether the effective
size varies across the different color bands, which 
provides information needed to understand the statistical distribution 
of stellar populations in galaxies for a global sample.  

The third quantity we study is the concentration 
index of the light distribution of each galaxy. 
This parameter is known to correlate with the morphological
type (Morgan 1958); 
earlier type galaxies have light profiles more concentrated toward
the center. There is some evidence that this parameter 
can be used for automated morphological classification 
(Doi et al. 1993; Abraham et al. 1994).
Specifically, we study the performance of automated classification
schemes when the concentration
parameter is used to classify galaxies into early and late types.

We describe briefly the data and the photometric catalog in \S 2.
In \S 3 our morphological classification and the resulting catalog 
are described.
Color indices are studied in \S4. 
In \S 5 the distribution of the scale lengths of galaxies is discussed, 
and in \S 6 the concentration index is studied. 
Conclusions are given in \S 7.  

%------------------------------------------------------------------

\section{Observation and photometric catalog}

The SDSS (York et al. 2000) is carried out 
using a wide-field 2.5m telescope, 
a large format imaging camera (Gunn et al. 1998), 
two fiber-fed double spectrographs, 
and an auxiliary 0.5m telescope for photometric calibration.
The sky is imaged with thirty photometric CCDs, arranged in six 
columns of five rows, each row corresponding to a different passband.  
The data are taken in time-delay-and-integrate mode at the sidereal
rate along great circles on the sky, yielding a {\em strip} consisting
of six very long and narrow {\em scanlines}, each $13^\prime.5$ wide.
The effective exposure time is 54.1 sec. 
The scanlines of a given strip do not overlap, but observing 
a second strip offset from the first strip by about $12^\prime.8$ gives
a filled {\em stripe} $2.^\circ5$ wide. 
The pixels are $0''.396$ square on a side.
The  0.5m Photometric Telescope observes standard stars to determine the
atmospheric extinction on each night and ties the standard stars
to objects observed with the 2.5m survey telescope.
SDSS imaging data taken in five passbands, 
$u',~g',~r',~i'$ and $z'$ (Fukugita et al. 1996), 
are processed with the photometric pipeline 
(hereafter {\it Photo}; Lupton et al. unpublished) specifically written 
for SDSS imaging data processing.

The galaxies used in this paper are taken from observations
of the Northern Equatorial stripe on 19 March, 1999 (SDSS Run 745),
supplemented with observations on
20 March, 1999 (Run 752) and 21 March, 1999 (Run 756).
These imaging data were taken before the commissioning
of the current Photometric Telescope. We calibrated the data by
observing secondary patches in the survey area using a (now
decommissioned) 61 cm telescope at the observatory site, and by
observing primary standard stars using US Naval Observatory's 
40-inch telescope, with filter and CCD characteristics nominally 
identical to those at the SDSS Photometric Telescope. 
Since the transformation from the primary standard stars 
to the objects observed with the SDSS 2.5m telescope 
has not yet been fully defined, we expect photometric errors of
$\approx\pm$0.05 mag in each band with respect to the proper SDSS
photometric system. 
Thus, in this paper, we will denote our magnitudes as 
$u^*$, $g^*$, $r^*$, $i^*$, and $z^*$, 
to emphasize the preliminary nature of our photometric calibration, 
rather than the notation $u'$, $g'$, $r'$, $i'$, and $z'$ 
that will be used for the final SDSS photometric system.  
However, we will use the latter notation to refer to the SDSS 
photometric passbands themselves. 
The magnitudes we use are based on the AB$_{95}$ 
system (Fukugita et al. 1996).

The galaxies we study in this paper are taken from Run 745,
which ranges from RA=$10^{\rm h} 41^{\rm m}$ to 
$16^{\rm h} 41^{\rm m}$   
covering effectively 127 square degrees in total.
We take all 217 galaxies brighter than $m_P(g^*) = 16$ mag
contained in Run 745 for morphological classification by eye
(here $m_P(g^*)$ stands for the Petrosian magnitude in the $g'$ band
as defined below).
Additionally, we take all 160
galaxies with $16 \le m_p(g^*) < 17$ in the RA range 
$10^{\rm h} 41^{\rm m}$ to $11^{\rm h} 41^{\rm m}$, and 107 galaxies
with $17 \le m_p(g^*) < 18$ in the RA range  $10^{\rm h} 41^{\rm m}$
to $10^{\rm h} 44.5^{\rm m}$. This makes the size of our 
morphologically classified sample to be 484 galaxies in total.
Our sample is not homogeneous with respect to brightness, but otherwise
we do not expect any other selection biases.
Those galaxies which fall on the edge of the fields are not taken
in our sample. The $g'$ band is used in order to facilitate 
comparison of our morphological classification with the work 
done in the past, which has been almost exclusively performed 
using the blue band image. 

After we carried out morphological classification using Run 745, 
an improved photometric data processing 
was done for Runs 752 and 756, the regions of which
overlap with that of Run 745. Therefore, we identified galaxies in our
sample with those imaged by Runs 752 and 756 and developed a photometric
catalog from these two runs\footnote{We used {\it Photo} 
version $5\_0\_3$ of late 1999. 
The most important improvement that concerns us here 
is in the accuracy of fluxes of deblended objects.}. 
Unfortunately, the entire region
of Run 745 is not covered with Runs 752 and 756, 
and this leads us to drop 28 galaxies. Hence we work with
456 galaxies in the analysis given in this paper.
Details of photometry of galaxies in Runs 752 and 756 are
discussed in Yasuda et al. (2000).

While many photometric parameters are measured in {\it Photo}, we are
concerned in this paper only with three classes of parameters:

\vspace{10pt}
\noindent
(1) {\it Petrosian radius in the five bands:} 
$r_P(u')$, $r_P(g')$, $r_P(r')$, $r_P(i')$, $r_P(z')$

The Petrosian radius $r_P$ is defined as the parameter that satisfies
\begin{equation}
\eta = { I(r_P) \over{2\pi \int^{r_P}_0 I(r)rdr/(\pi r_P^2) }} .
\end{equation}
\noindent
In practice, this implicit equation for $r_P$ is replaced with
\begin{equation}
\eta = { 2\pi \int^{1.25r_P}_{0.8r_P} I(r)rdr
               / [ \pi((1.25r_P)^2-(0.8r_P)^2)]
       \over{ 2\pi \int I(r)rdr/(\pi r_P^2) }},
\end{equation}
\noindent
where $I(r)$ is the surface brightness profile of the object.
We adopt $\eta=0.2$.
We measure the Petrosian radius in each band, but we use
only the radius measured in the $r'$ band to calculate the
Petrosian flux (or Petrosian magnitude) for all color bands,
so that the aperture for the Petrosian flux is common to the 
five passbands.

\vspace{10pt}
\noindent
(2) {\it Petrosian magnitude in the five passbands:} 
$m_P(u')$, $m_P(g')$, $m_P(r')$, $m_P(i')$, $m_P(z')$

The Petrosian flux is defined as
\begin{equation}
F_P = 2 \pi \int^{k r_P}_0 I(r)rdr,
\end{equation}
\noindent
where $k$ is set equal to 2 
and $r_P$ is the $r'$ Petrosian radius. 
For further detailed discussion of the
Petrosian flux, see Yasuda et al. (2000).

\vspace{10pt}
\noindent
(3) {\it Petrosian half light radius and  Petrosian $90\%$ light 
radius:} 
$r_{50}(\lambda_i)$, $r_{90}(\lambda_i)$

$r_{50}$ and $r_{90}$ are defined, respectively,
so that the fluxes over the aperture with these  
radii are  
$50\%$ and $90\%$ of the Petrosian flux 
in the passband $\lambda_i=u',g',r',i'$, and $z'$.

To give an idea about the basic properties of our sample, 
we show the distribution of $g'$-band Petrosian magnitudes 
in Figure 1, and those of $r_P(r')$, $r_{50}(r')$, 
and $r_{90}(r')$ in Figure 2.
Magnitudes are corrected for Galactic extinction 
using the dust distribution estimated by Schlegel, 
Finkbeiner, \& Davis (1998) with the ratio of total to
selective extinction $R_{g'}=3.79$ and $R_{r'}=2.75$.
When necessary we use $R_{u'}=5.2$, $R_{i'}=2.09$,
and $R_{z'}=1.48$ for the other color bands.

\begin{figure}[t]
\epsscale{1.2}
\plotone{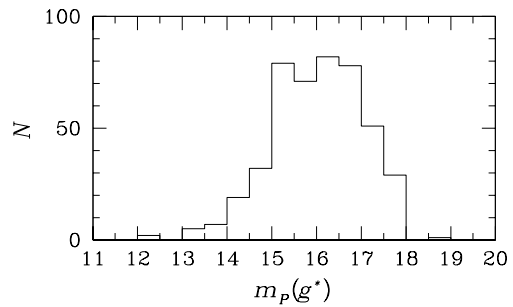}
\caption
{Distribution of Petrosian magnitudes in the $g'$ band for
456 galaxies analyzed in this paper.
Magnitudes are corrected for Galactic extinction.
\label{fig1}}
\end{figure}

\begin{figure}[t]
\plotone{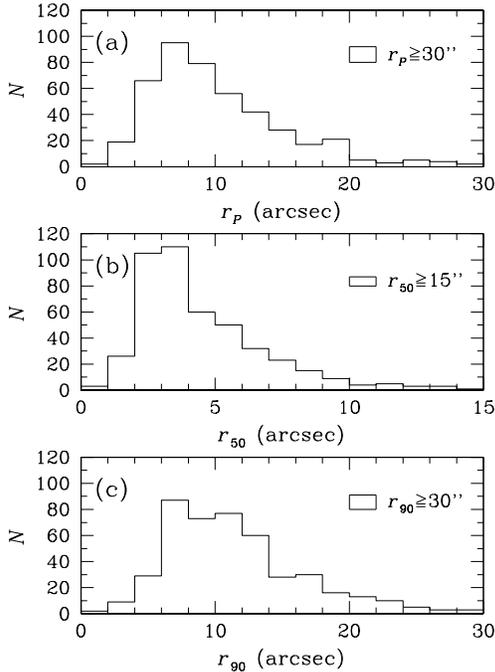}
\caption
{Distributions of Petrosian radii, Petrosian half light radii,
and Petrosian $90\%$ light radii of the 456 galaxies
measured for $r'$-band images.
\label{fig2}}
\end{figure}

%------------------------------------------------------------------

\section{Morphological classification}

The galaxies in our sample 
are classified into 7 morphological classes, 
$T=0$ (corresponding to E in the Hubble type, or $-5$ 
in RC3 type), 
1 (S0, $T_{\rm RC3}=-3$ to $-1$), 
2 (Sa, $T_{\rm RC3}=1$), 
3 (Sb, $T_{\rm RC3}=3$), 
4 (Sc, $T_{\rm RC3}=5$), 
5 (Sdm, $T_{\rm RC3}=8$), 
and 6 (Im, $T_{\rm RC3}=10$), 
where $T_{\rm RC3}$ refers to the type index defined in the
{\it Third Reference Catalogue of 
Bright Galaxies} (de Vaucouleurs et al. 1991, hereafter RC3).
This classification is coarser than the standard RC3 types, but we
consider that our classification is sufficient for most purposes
for galaxy science.
Four of the present authors (MH, TI, SO, and MS) 
independently classified all galaxies by eye, 
by comparing the image of each galaxy displayed on the SAOimage viewer
%at an appropriate image size and contrast,
with template galaxies of Frei \& Gunn (1994) which are given 
morphological types by RC3.
The templates of Frei \& Gunn cover from E to Im.
When one cannot assign a morphology to a galaxy, an index 
of $-1$ is given.
For each galaxy, we adopt a median of four classifiers
as the final value of $T$ (a half integer means that the median falls
in the middle).
In finding a median we ignore $T=-1$ when one or two classifiers give
this index (no galaxy was given $-1$ by more than two classifiers).
In this way some galaxies are given half integer values of $T$.
The catalog will be published elsewhere 
(Fukugita et al., unpublished).

We have identified the 54 galaxies listed in RC3 and in Run 745.
Figure 3 compares our morphology $T({\rm ours})$ with that given in 
RC3, $T_{\rm RC3}$, for the 54 galaxies.
The size of the symbol is set so that its 
area is proportional to the number of galaxies in the grid.
The curve connects points where our morphology matches 
with RC3 morphology.
A tight correlation is seen between the two classifications.
For a given $T({\rm ours})$ the average discrepancy 
in the two classifications (in units of $T_{\rm RC3}$) is 
$\langle|\Delta T_{\rm RC3}|\rangle=1.6$ with the rms scatter of 1.8.
Naim et al. (1995) compared experts' classifications and
concluded that the rms scatter among classifiers is
$1.8$. Somewhat larger discrepancy between our classification 
and that given in RC3 may be expected, owing to
the fact that we have estimated morphology from CCD images for
which the contrast and grey scale are not fixed, 
whereas RC3 classification
is based on photographic materials. On the other hand, Naim et al. used 
the identical prints for test classifications. 
In view of these considerations we take
our classification to be acceptable. 
In specific details, 
our classification tends to classify some S0 galaxies as E, 
and classify spiral galaxies into somewhat later types.

\begin{figure}[t]
\plotone{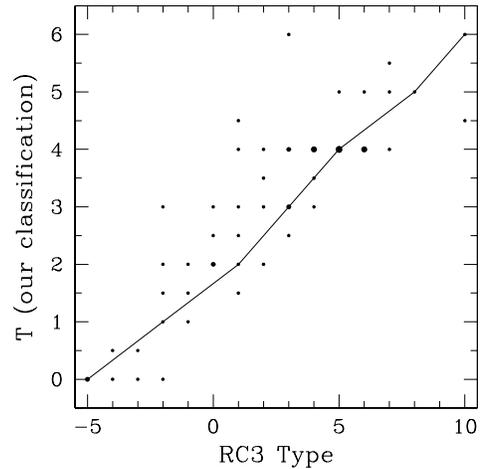}
\caption
{Morphology determined by our eye classification
plotted against that given in RC3 for 54 common galaxies.
The solid line connects points where our morphology matches
with RC3 morpholgy.
\label{fig3}}
\end{figure}

\begin{figure}[t]
\plotone{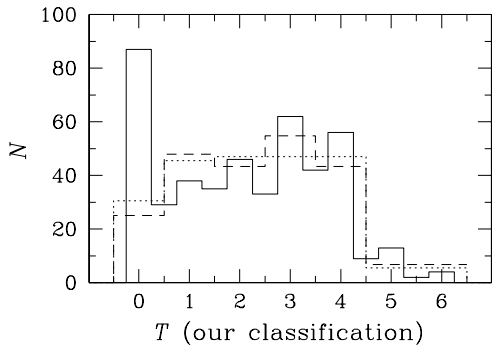}
\caption
{Distribution of morphological types for the 456 galaxies
(solid line).
The dotted line and dashed line indicate the type mixes
given in Loveday (1996) and Fukugita et al. (1998), respectively.
\label{fig4}}
\end{figure}

Figure 4 shows the morphological type distribution of 
galaxies in our sample. The relative fractions of types
given in the literature (Fukugita, Hogan, \& Peebles 1998; 
Loveday 1996) obtained in the $B$ or $b_J$ band
are also plotted for comparison 
(normalized to the total number of galaxies in our sample). 
Loveday does not classify
spiral galaxies into subclasses; 
so his numbers are distributed evenly over classes 2, 3, and 4.
In our sample the number of galaxies with $T=0$ (i.e., E) is 
significantly higher than that of $T=1$ (S0) galaxies, in
contrast to the other results, implying
that our classifiers set the division criterion
of E and S0 to be more to the S0 side than in other work 
\footnote{In a later work, we tried to reclassify the sample 
by resetting the division criterion
in a way that the frequency of E:S0 agrees with 
more traditional values in the blue band,
but we do not adopt this reclassification in the present paper.}.
Taking $T < 1.5$ as early type (E and S0) galaxies, the frequency of
the early type galaxies in our sample $0.34 =154/456$ is consistent
with what has been known from the work in the blue bands. 
In the following analysis we mostly take E and S0 as a single class.
%We also note that the frequency of Im galaxies is smaller than that
%often given in literature.
To conclude this section, we quote the frequency of morphological types:
\begin{equation}
{\rm E+S0:Sa:Sb:Sc:Sdm+Im = 0.34: 0.18 : 0.21 : 0.21 : 0.06}.
\end{equation}
We do not detect any significant systematic change of morphological 
compositions with magnitude bins. So we regard this fraction to
represent a fair value for field galaxies brighter than $r^*=18$ mag.

%------------------------------------------------------------------

\section{Color indices}

\subsection{Color indices as a function of morphology}

In Figure 5 we plot (a) $u^*-g^*$, (b) $g^*-r^*$,
(c) $r^*-i^*$, and (d) $i^*-z^*$ colors against $T$.
The size of the symbol is again set so that its 
area is proportional to the number of galaxies falling
in the grid. The mean and rms of the color distributions 
are calculated in Table 1, and represented in 
Figure 6 (a)-(d)\footnote{When calculating the mean, 
we dropped galaxies having too
unrealistic colors; their colors are likely to be ascribed to errors 
due to unsuccessful deblending processes.}.
As noted in section 2, 
colors of each galaxy are measured for a common aperture 
which is defined as $2 \times r_P(r^*)$. The aperture is
reasonably large, so that our color indices are regarded as
representing nearly those for the total flux. 
We correct all colors for Galactic reddening.
The open squares connected by lines in Figures 5 and 6 
are colors calculated 
convolving Kennicutt's (1992) spectrophotometric
atlas of nearby galaxies with the responses of the SDSS 
photometric system to give broad band colors
(Fukugita, Shimasaku, \& Ichikawa 1995; hereafter FSI).

\begin{figure}[t]
\plotone{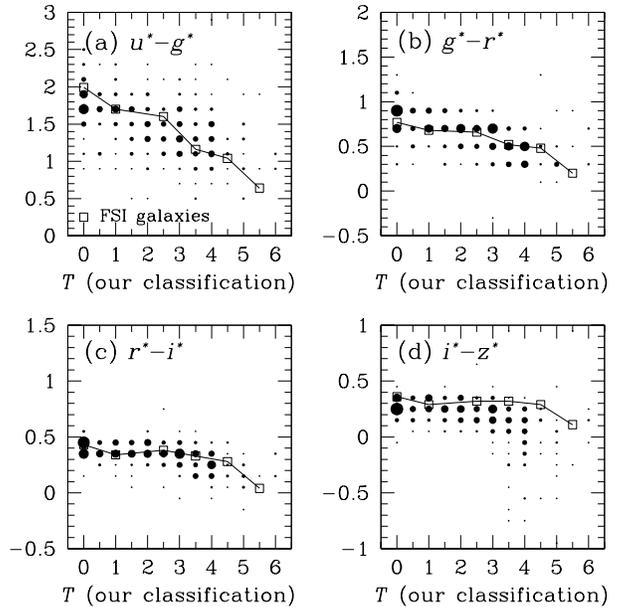}
\caption
{Colors of our galaxies plotted against morphology.
Panels (a), (b), (c), and (d) are for $u'-g'$, $g'-r'$,
$r'-i'$, and $i'-z'$, respectively.
All colors have been corrected for Galactic reddening.
The area of each filled circle is proportional to the number
of galaxies falling into the grid.
The open squares connected with solid lines indicate
colors calculated (FSI) from Kennicutt's spectrophotometric atlas
of nearby galaxies.
\label{fig5}}
\end{figure}

\begin{figure}[t]
\plotone{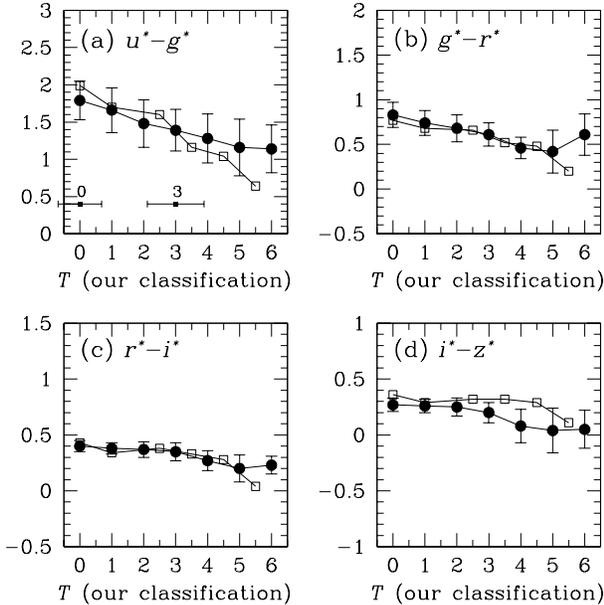}
\caption
{Mean and rms colors of dereddened galaxies plotted against morphology.
Panels (a), (b), (c), and (d) are for $u'-g'$, $g'-r'$,
$r'-i'$, and $i'-z'$, respectively.
The open squares connected with solid lines indicate
synthetic calculation of FSI.
The two filled squares with error bars in (a) 
show the errors in $T({\rm ours})$ measurements at $T=0$ and 3 
estimated from the difference between our classification and that 
of RC3.
\label{fig6}}
\end{figure}

We indicate in Fig.6(a){\hspace{5pt}} representative uncertainties 
in our $T$ measurements for early and late types  
by horizontal bars (the estimate at $T=0$ and 3) 
which are 
estimated from the difference between our classification and that 
of RC3 shown in Fig.3. \footnote{
The uncertainty we plotted here is  
the quadratic sum of 
the random and systematic errors around the RC3 type
at a given $T$.}

We observe (more clearly in Figure 6) 
that the data are distributed fairly close to the
FSI calculations for (a) $u^*-g^*$, (b) $g^*-r^*$, and
(c) $r^*-i^*$ colors.  For $g^*-r^*$ the difference between
the mean of the data and the calculation is less than 0.05 
except for Im (a 0.3 mag discrepancy), for which
the synthesis calculation uses the bluest known galaxies 
(NGC4449 and NGC 4485) to define the envelope of the bluest color.
The next largest discrepancy in $g^*-r^*$ color 
is seen for elliptical galaxies:
the data are redder than the synthesis calculation by 0.05 mag.
This level of discrepancy, however, is seen for $B-V$ color of
elliptical galaxies,
either with Kennicutt's spectrophotometric data 
or with the spectroscopic energy distribution (SED) of Coleman, Wu, 
\& Weedman (1980, hereafter CWW), when compared with the broad band
color: the template SED of elliptical galaxies is not sufficiently red
to give $B-V\simeq 1$ obtained by broad band photometry, 
as noted by FSI. 

For $r^*-i^*$ color the difference between the data and the FSI 
synthetic colors 
is less than 0.04, again except for Im galaxies. 
The disagreement between our data and the FSI calculations 
is somewhat larger
for $u^*-g^*$, but the scatter is wider for this color; we cannot
conclude that the disagreement is significant. We remark, however, that
the calculation being redder than observation by 0.2 mag for
E galaxies may be ascribed
to the use of metal rich giant elliptical galaxies in constructing the
composite SED. 
The $u^*-g^*$ index is most sensitive to the metal abundance. 
For Im galaxies 
the observed $u^*-g^*$ color is redder than the calculation 
because of the use of a blue template object, 
as we have seen for the Im type in $g^*-r^*$.

A significant deviation is seen for the $i^*-z^*$ color, 
for which the data are clearly bluer
than the synthetic calculation by 0.1 mag for early-type galaxies
and 0.2 for later-type galaxies. While the SED 
that covers the $z'$ band used in FSI relies on an extrapolation from 
the bluer wavelengths, we
cannot blame this disagreement simply on an incorrect extrapolation,
as we discuss in section 4.2.

We present in Figure 7 color-color diagrams 
(a) $g^*-r^*$ vs. $u^*-g^*$, (b) $r^*-i^*$ vs. $g^*-r^*$, 
and (c) $i^*-z^*$ vs. $r^*-i^*$. 
The dots represent each galaxy, 
and the curves are predictions from synthetic
calculations of FSI (filled squares), CWW SEDs
extended to the near infrared by Neugebauer (personal communication)
(filled triangles), both of which use empirical SED of galaxies, 
Kodama \& Arimoto's (1997) stellar population synthesis
model (open squares), 
and GISSEL of Bruzual \& Charlot's 
(1996, unpublished; see also {\it idem} 1993) 
stellar population synthesis model (open triangles).
The convergence of the four synthesis calculations is very good for the 
$g^*-r^*$ vs. $u^*-g^*$ plot. 
A trend is visible that the SDSS photometry
gives a somewhat bluer $u^*-g^*$ (or slightly 
redder $g^*-r^*$) for
early type galaxies than all synthetic calculations.
The agreement among synthetic calculations is still fairly good for
the  $r^*-i^*$ vs. $g^*-r^*$ plot; the SDSS data also closely follow
the synthetic calculations. In the $i^*-z^*$ vs. $r^*-i^*$ plot,
however, the data are clearly shifted by 0.1$-$0.2 mag downward
(bluer $i^*-z^*$) compared with all four synthetic curves,
which mutually agree very well.
\footnote{
This does not necessarily mean that
model SED's are correct, since they are not well constrained
beyond $i'$ pass band.} 
The $i'-z'$ color of the SDSS galaxies 
is deviated from all model calculations.

\setcounter{figure}{6}
\begin{figure}[t]
\plotone{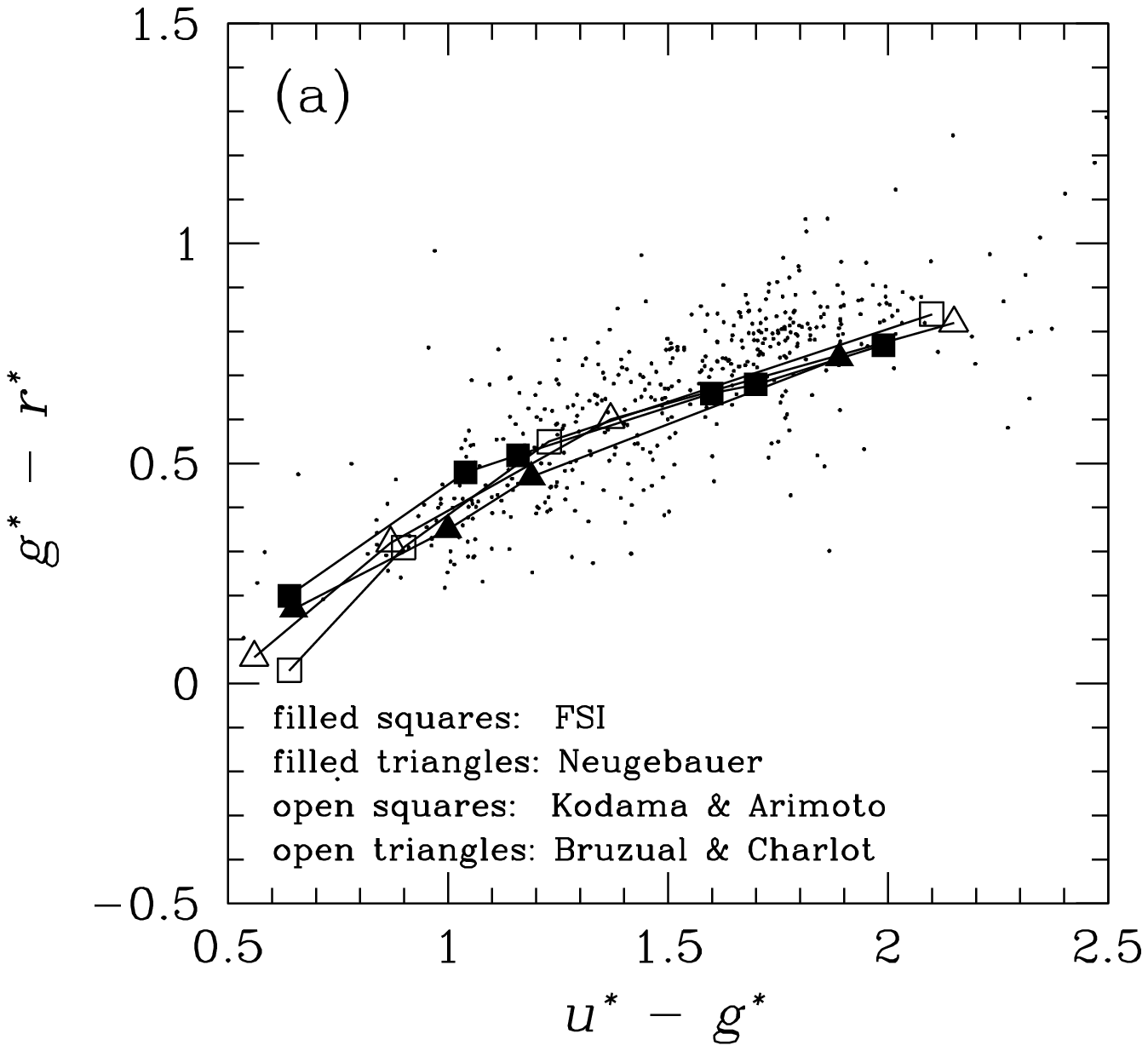}
\caption
{(a) Color-color distribution of our galaxies:
$u^*-g^*$ vs $g^*-r^*$.
The dots represent each galaxy,
and the curves are predictions from synthetic
calculations of FSI (filled squares),
CWW SEDs
extended to the near infrared by Neugebauer (personal communication)
(filled triangles),
Kodama \& Arimoto's (1997) stellar population synthesis
model (open square),
and GISSEL of Bruzual \& Charlot's (1993)
stellar population synthesis model (open triangles).
\label{fig7}}
\end{figure}

\setcounter{figure}{6}
\begin{figure}[t]
\plotone{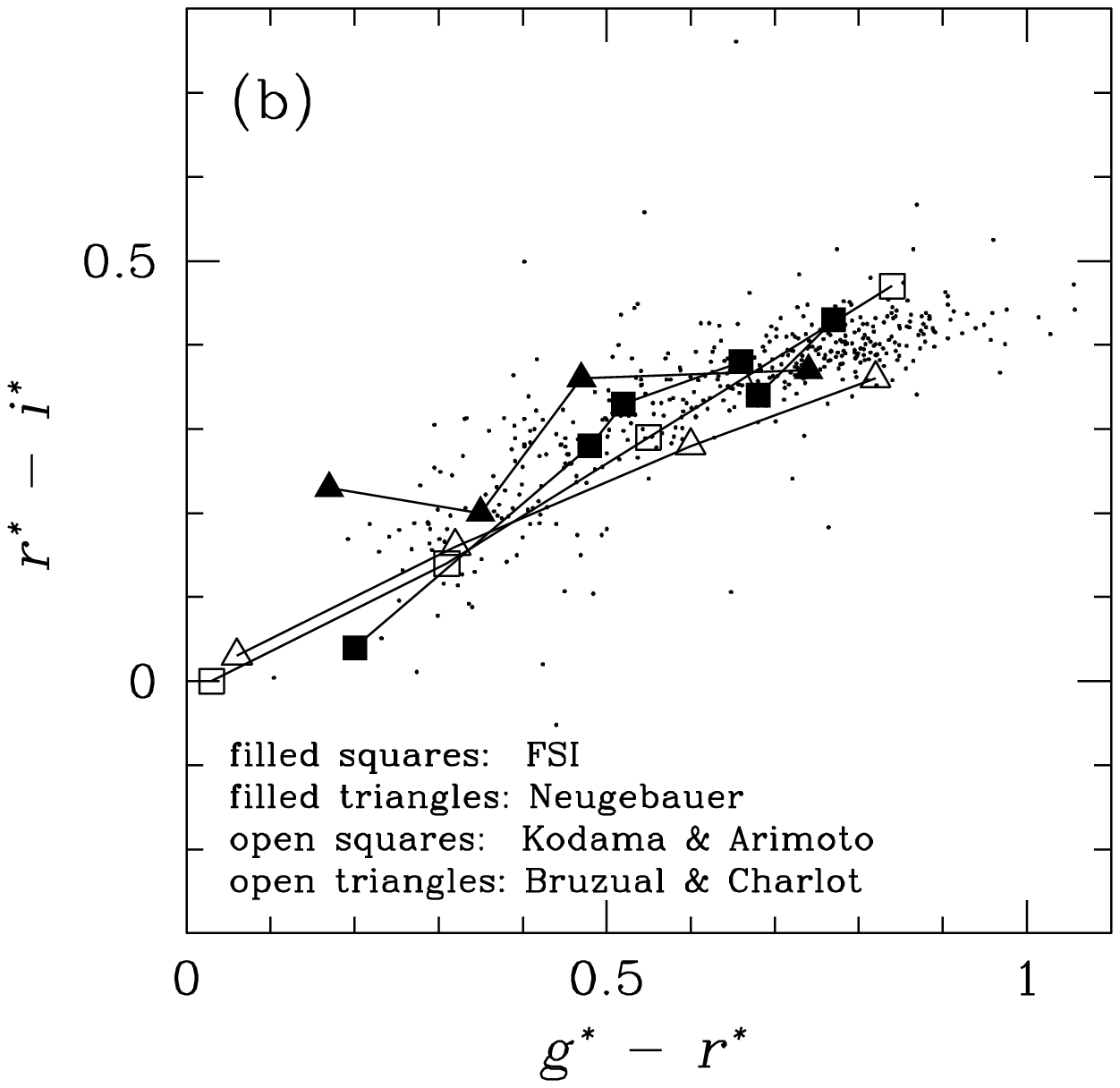}
\caption
{(b) Same as (a), but for $g^*-r^*$ vs $r^*-i^*$.
\label{fig7}}
\end{figure}

\setcounter{figure}{6}
\begin{figure}[t]
\plotone{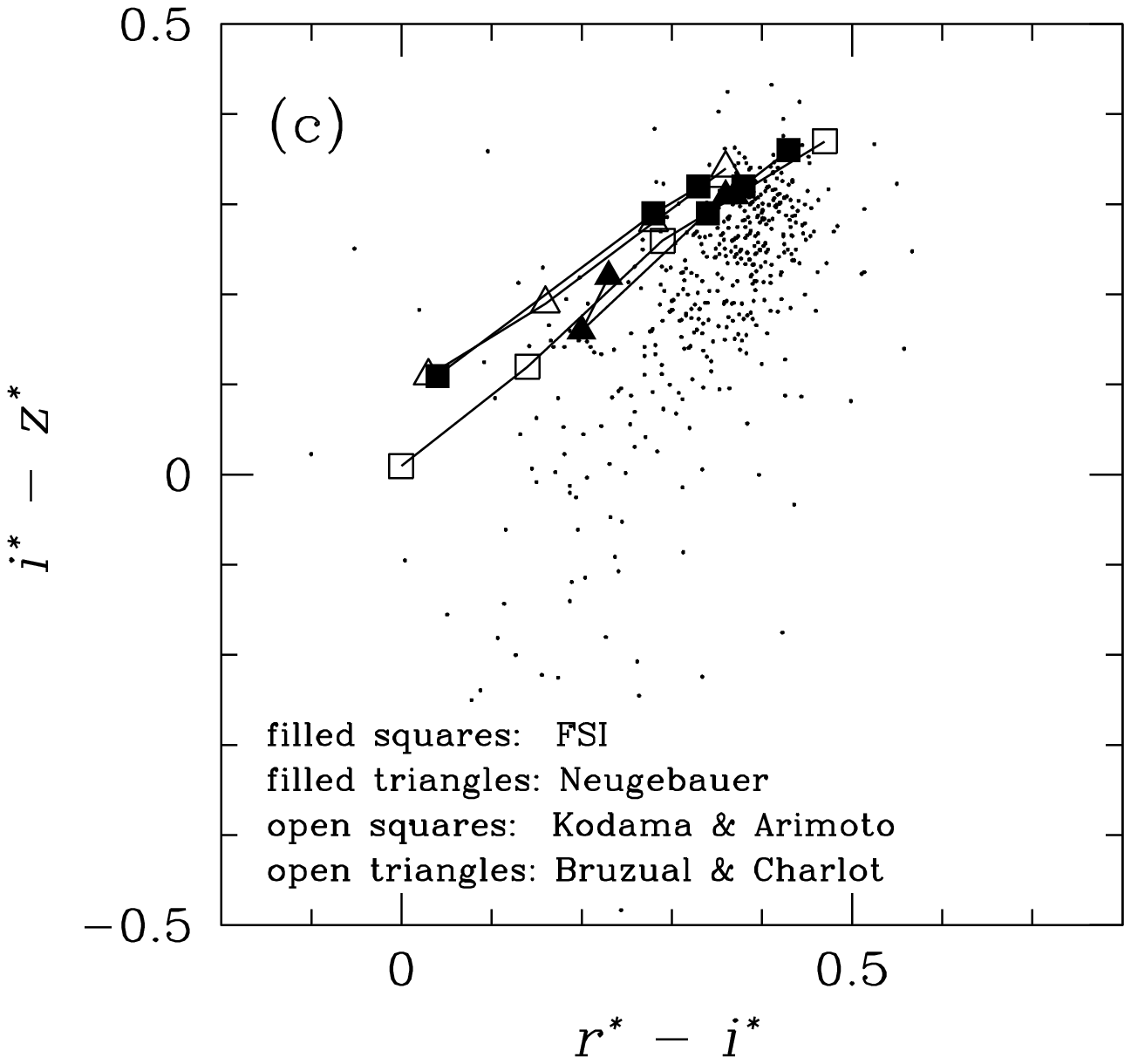}
\caption
{(c) Same as (a), but for $r^*-i^*$ vs $i^*-z^*$.
\label{fig7}}
\end{figure}

\subsection{Comparison with other multicolor broadband photometry}

Buta \& Williams (1995, hereafter BW) present $VR_CI_C$ colors 
for 501 nearby galaxies
classified into morphology according to the RC3 system.
$UBV$ colors for these galaxies are given in RC3. 
In this subsection, we compare SDSS photometry with the
BW (and RC3) catalog to examine the consistency among 
broad band photometries.  
Color transformations from the $UBVR_CI_C$ system to the SDSS system are
carried out using the transformation laws given in FSI. 
The RC3 morphology indices are also converted into our system.
We adopt BW's total color to compare with our Petrosian color.
Figure 8(a)-(d) gives plots for the BW sample, 
similar to the plots in Figure 5(a)-(d).
We remark that the BW sample is not homogeneous with respect to
morphology, but is highly biased towards early type galaxies. 
We expect, however, that this does not produce any biases to
our color analysis.
We overlay the data of our sample (Fig.5; we omit $T=0.5$ and $1.5$ 
data since no galaxy is classified as these types in the BW sample)
by open circles, 
with a horizontal offset of $+0.2$ for clarity.

\setcounter{figure}{7}
\begin{figure}[t]
\plotone{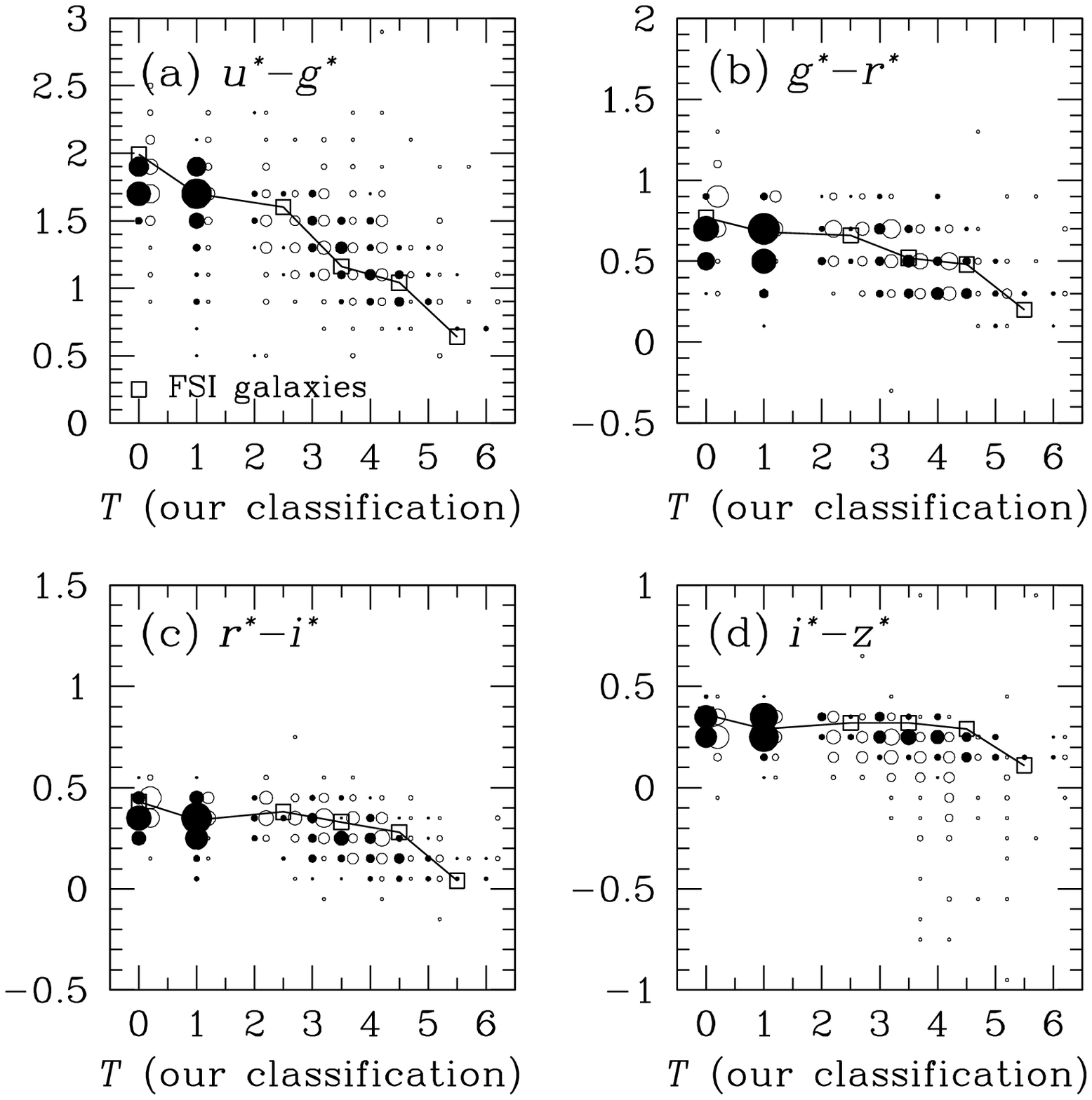}
\caption
{Same as Fig.5, but for Buta \& Williams' (1995) galaxies.
Overlaid are the plots for our sample (Fig.5) by open circles  
with a horizontal offset of $+0.2$ for clarity.
\label{fig8}}
\end{figure}

A comparison of Figure 8 with 5 shows that 
the $u'-g'$, $g'-r'$, and $r'-i'$ distributions of the BW galaxies
are grossly consistent with those for SDSS photometry.
The agreement of the $u'-g'$ colors between the two samples is
particularly good, although a somewhat smaller scatter for the
early type galaxies of the BW sample may be ascribed to its selection
effect that the BW sample is not homogeneous.  
The $g'-r'$ and $r'-i'$ colors for early type galaxies
are slightly bluer for the BW sample. 
The appreciable discrepancy found in the comparison of the two samples
is with $i'-z'$ colors.
$i'-z'$ colors of the SDSS sample are systematically 
bluer than those of BW by $\sim 0.1-0.2$ mag.
This trend is similar to what was found between the FSI calculation 
and the SDSS measures of galaxies in the previous subsection.

This discrepancy in $i'-z'$ color
leads us to suspect that there may be a systematic error by 
$0.1-0.2$ mag in $z'$ band photometry,
but the real reason is not clear to us.
The preliminary work for the SDSS primary standard stars 
(Smith et al. 2001)
shows that all SDSS color indices relative to the
Johnson-Morgan-Cousins photometric system differ 
from a synthetic calculation
using Gunn-Stryker stars (Gunn \& Stryker 1983) 
by no more than 0.03 mag.
We have recently recognized some unexpected
shifts of the filter response function 
in the 2.5 m camera system due to the evaporation of water vapor
from silicon layer of the coating surface when the filters are
soaked in a vacuum chamber, but we estimate the effect to be 
smaller than 0.015 mag for all color bands. 
The difference in the response between the front-illuminated thick
CCD device used in the 2.5m camera and the back-illuminated thin device
for the Photometric Telescope (which defines the SDSS magnitude) does
not cause an offset of more than 0.01 magnitude. 
The sky brightness and its time variation are the highest 
in the $z'$ band, so 
errors in the sky subtraction in $z'$ band 
could affect photometry of galaxies.
However, it is not easy to estimate photometric errors 
due to the sky subtraction errors, 
since they affect galaxy photometry 
in a complicated way 
depending on sizes and surface brightnesses of galaxies. 
At present, we have not found clear evidence that they 
account for the discrepancy in $i'-z'$ color. 
It should be, however, worth noting that 
the observed amount of the discrepancy in $i'-z'$ 
does not depend either on the brightness (and size) of galaxies 
nor on the position of galaxies on a strip, 
ie, when they were imaged.
The effect of 0.1-0.2
mag seems a bit too large to be ascribed to the uncertainties in the
present photometric system, but we cannot find the reason for this
discrepancy at the moment. 

%------------------------------------------------------------------

\section{Scale lengths}

In this section, we study the relation between apparent 
size and apparent brightness of galaxies. We expect different 
relations for different morphological types. The relation may
also depend on the color band used, if star forming 
regions or metallicity distributions are not homogeneous throughout
the galaxy. The information on the radius-brightness relation  
is also needed to construct a realistic simulation of galaxies. 
We additionally examine whether the relation derived from
SDSS photometry is consistent with our knowledge of the galaxy
sizes we obtained from other studies of local galaxies.
The consistency would allow us to supplement the SDSS data 
with the results from other, more detailed studies, 
such as which have not yet been done with SDSS.

We consider $r_{50}$ estimated in the five color bands as a 
measure for the galaxy size. Note that
$r_{50}$ is approximately the conventionally defined 
effective radius (half light radius).
We adopt Petrosian magnitudes, $m_P$, also in the five color bands
to represent apparent brightness.

Four panels in Figure 9 show the plot of $r_{50}(\lambda_i)$ 
($\lambda_i=u',~g',~r'$, and $i'$) as a function of apparent brightness 
in the $r'$ passbands.
We divide our galaxy sample into three morphological 
classes on the basis of $T$ values:
the open circles are early type galaxies with $T < 1.5$ 
(E and S0 galaxies), the filled circles denote $1.5 \le T <4$
(S0/a to Sbc), and the crosses are for late spiral galaxies with 
$T \ge 4$ (later than Sc).
The data are well represented by the curve
\begin{equation}
\log r_{50}(\lambda_i)'' 
  = -0.2 \left[ m_P(\lambda_j)-16~ {\rm mag} \right]+ a_{ij},
\end{equation}
\noindent
where $a_{ij}$ are constants that depend on $\lambda_i$ and 
$\lambda_j$. 
This relation holds for
any combination of the passbands $\lambda_i,\lambda_j$ with the 
parameters $a_{ij}$ (and their rms scatters) 
being obtained by least squares fits to the plots
for the three morphological classes.
These parameters are given in Table 2 for 
$\lambda_i=u',~g',~r',~i',~z'$, $\lambda_j=r'$, 
and for $\lambda_i=\lambda_j= u',~g',~r',~i',~z'$. 

\setcounter{figure}{8}
\begin{figure}[t]
\plotone{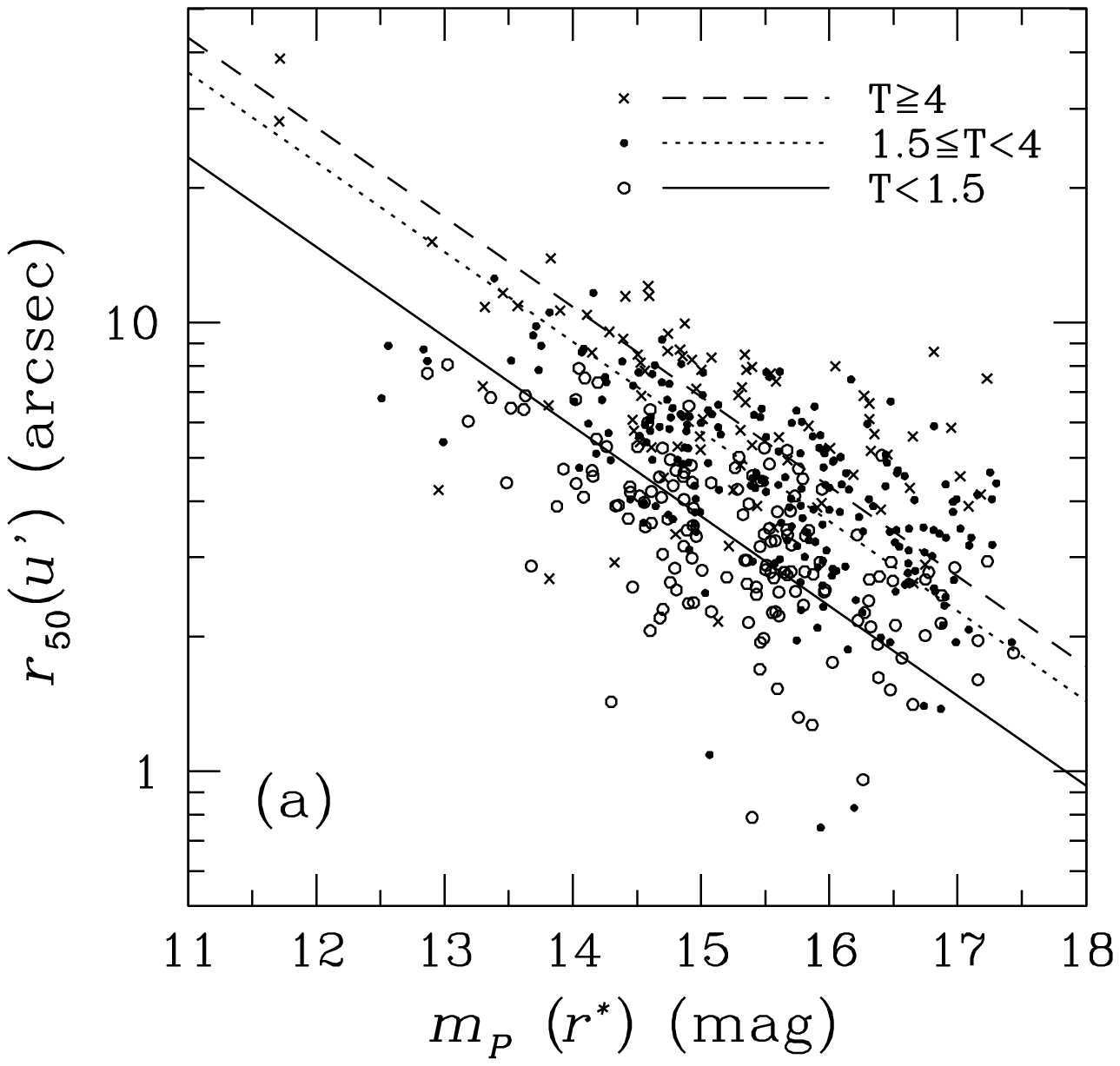}
\caption{(a) Half light Petrosian radii measured in $u'$,
plotted as a function of Petrosian magnitude in the $r'$ band.
Magnitudes have been corrected for Galactic extinction.
Open circles, filled circles, and crosses indicate
$T < 1.5$, $1.5 \le T < 4$, and $T \ge 4$, respectively.
The solid, dotted, and dashed lines in each panel show
the best fit of eq. (5) to the three morphological
classes of galaxies, respectively.
\label{fig9}}
\end{figure}

\setcounter{figure}{8}
\begin{figure}[t]
\plotone{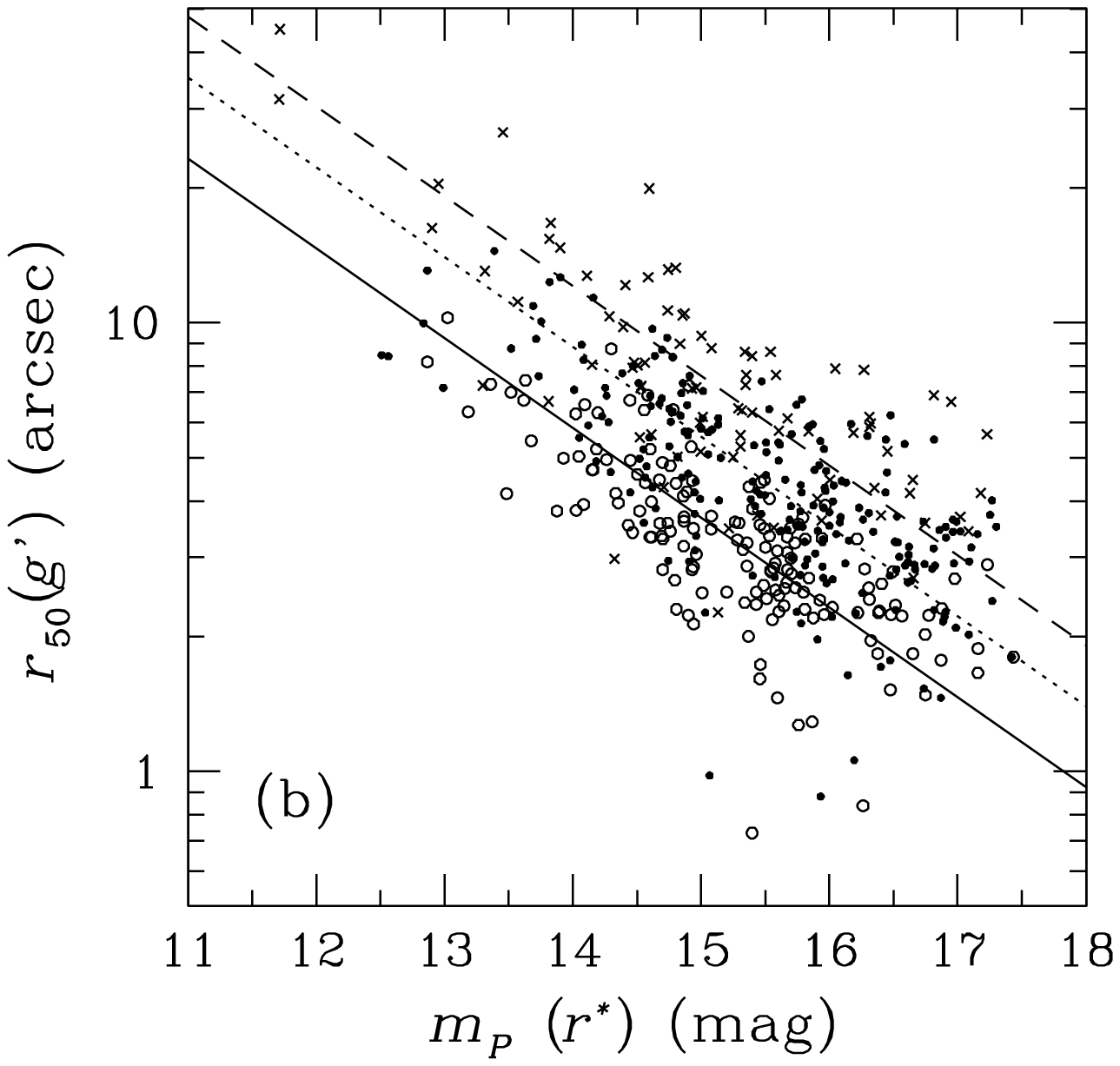}
\caption{(b) Same as (a), but for $g'$.
\label{fig9}}
\end{figure}

\setcounter{figure}{8}
\begin{figure}[t]
\plotone{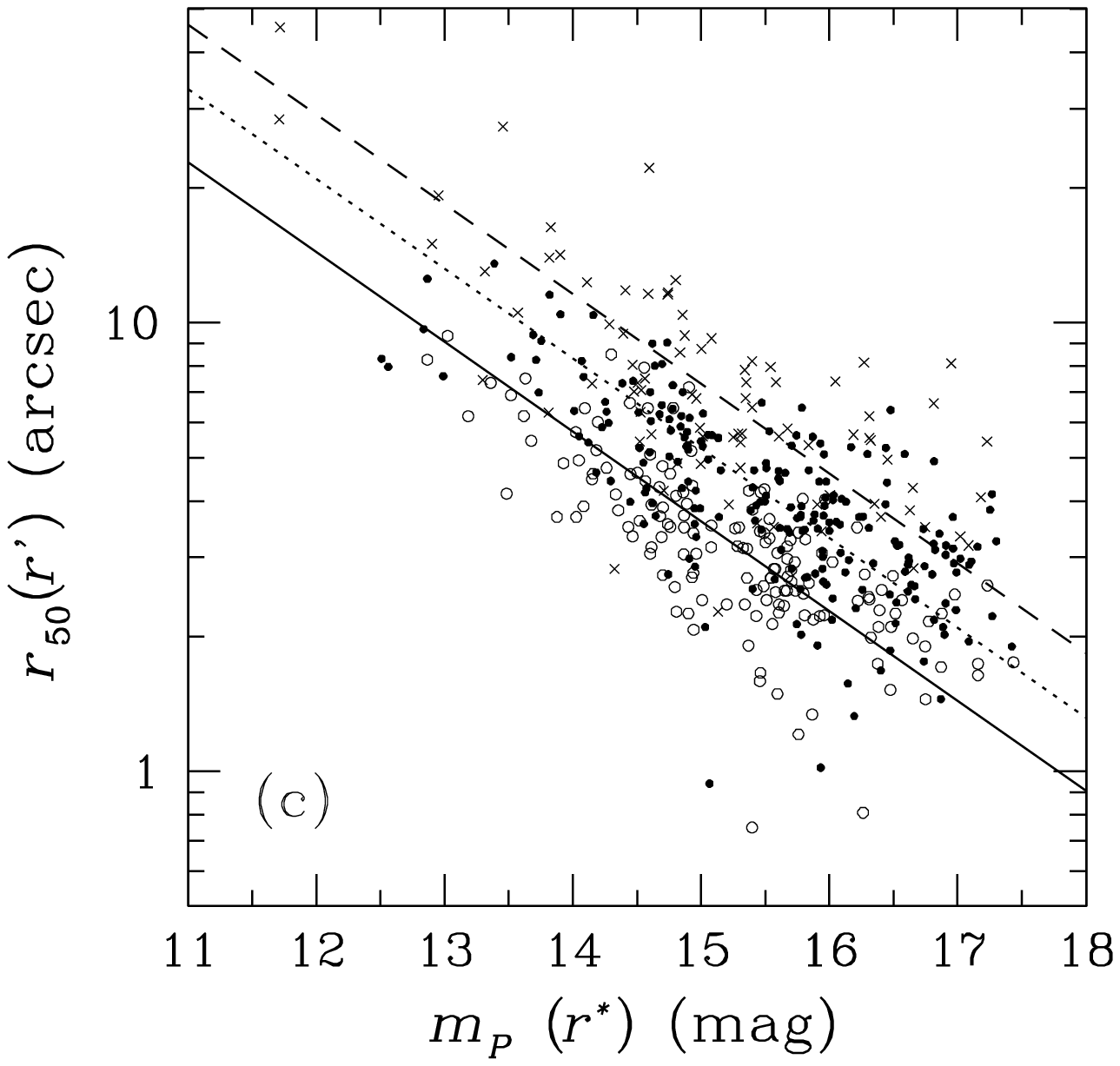}
\caption{(c) Same as (a), but for $r'$.
\label{fig9}}
\end{figure}

\setcounter{figure}{8}
\begin{figure}[t]
\plotone{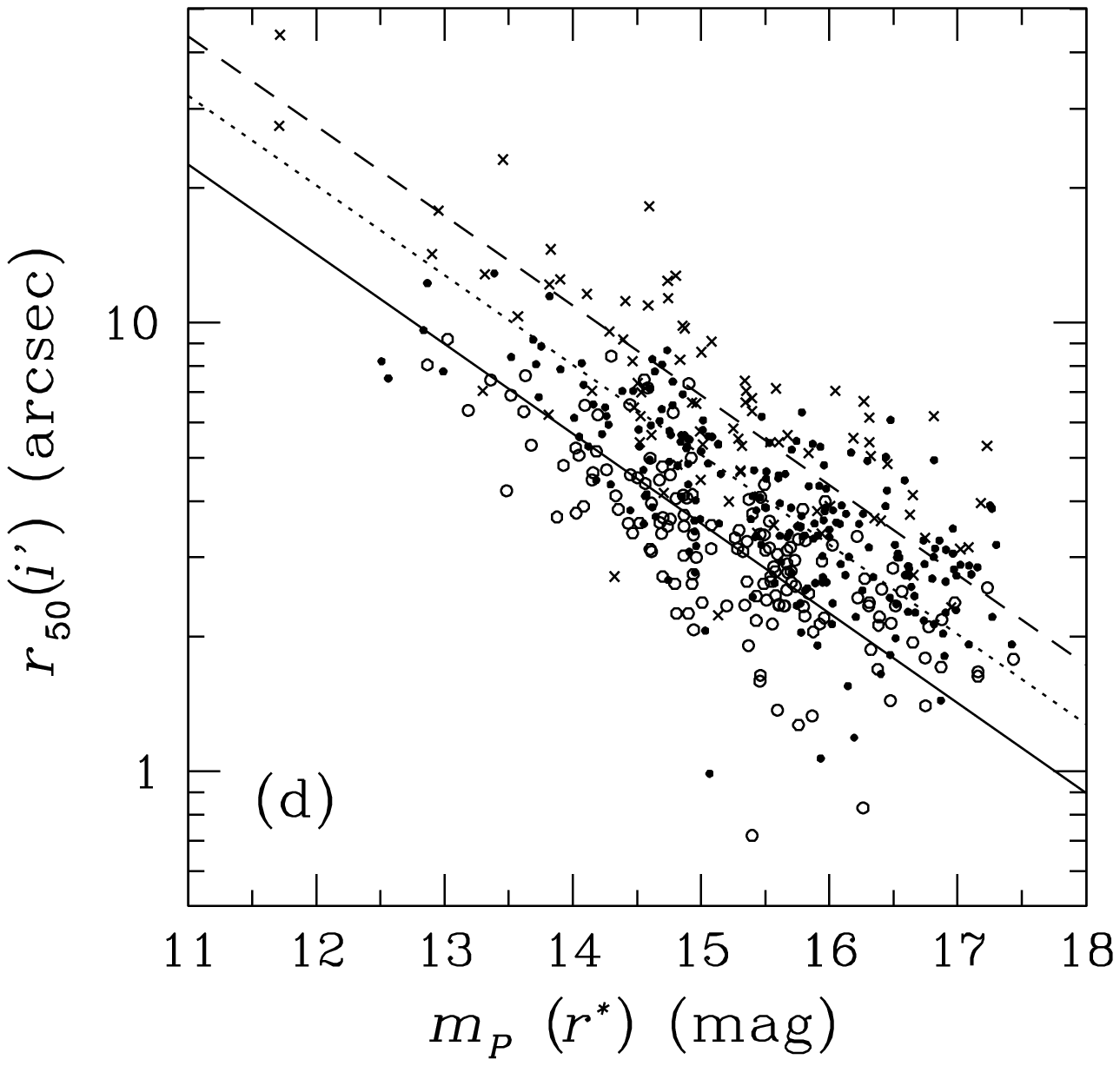}
\caption{(d) Same as (a), but for $i'$.
\label{fig9}}
\end{figure}

The plots in Figure 9 show that half light Petrosian radii $r_{50}$
are nearly independent of the color bands with which these
radii are calculated. Looking into details, however, we see 
the trend that the half light radius shrinks slightly from UV to 
redder bands for all morphological types (see rows 1-5 of Table 2). 
For early type galaxies we observe a 10\% decrease in the effective
radius continuously from the $u'$ to the $z'$ band. This
is interpreted as due to the effect of the color gradient of 
early type galaxies driven from the metallicity gradient.
For spiral galaxies the decrease is 20-25\%. This 
represents the star forming component being more extended 
towards outer parts of galaxies:
in other words an increasingly dominant importance of the 
bulge component in redder passbands.

The fitting formulae with $i=j$ 
give useful parameters (see rows 6-10 of Table 2) to 
construct simulations
of galaxies. 
The knowledge of these parameters is particularly important in
understanding the detection limit of galaxies and the performance of
star galaxy classification.

Let us now investigate the consistency of eq.(5) with the 
relation involving the effective radius known in the literature.
The most accurate study concerning the scale of galaxies is
presented by Kent (1985) on the basis of excellent Thuan-Gunn $r$ band
CCD images of nearby bright galaxies with known redshifts.
He derived the relations between intrinsic size 
and surface brightness separately for the bulge and disk components.
Using the transformation law of FSI, we convert his  
relations between the effective radius (kpc) and the effective surface 
brightness (mag arcsec$^{-2}$) to those in the SDSS $r'$ band as,
\begin{equation}
\mu_e = 2.5\log [r_e~({\rm kpc})] + 20.3 ({\rm mag~sec}^{-2})\ ,  
\hspace{20pt} {\rm (E,S0)}
\end{equation}
\begin{equation}
\mu_e = 2.5\log [r_e~({\rm kpc})] + 20.5  ({\rm mag~sec}^{-2})\ , 
\hspace{20pt} {\rm (disks)}
\end{equation}
\noindent
where 
we have assumed the Hubble constant to be 100 km s$^{-1}$ Mpc$^{-1}$.

When these relations are fitted to the SDSS data for $T<1.5$ (bulge
dominated) and $T \ge 4$ (disk dominated), we obtain for
the effective radius $r_e$ defined by 
$I(r)\propto\exp[-0.7289(r/r_e)^{1/4}]$
for the de Vaucouleurs profile and the scale length defined by
$I(r)\propto\exp(-r/h)$ for the
exponential disk,

\begin{equation}
r_e=2.56^{+0.81}_{-0.62} ~ {\rm kpc}\ ,~~~~  
({\rm de~Vaucouleurs~profile})
\label{eq:re-deV}
\end{equation}
\noindent
and
\begin{equation}
h=2.62^{+0.83}_{-0.63}~ {\rm kpc}\ ,~~~~  ({\rm exponential~disk})
\label{eq:re-exp}
\end{equation}
\noindent
where the errors indicate the $1\sigma$ range. Note that the average
luminosity of galaxies that contribute to galaxy number counts
in the bright magnitude is very close to $L^*$ 
(within $\pm$0.1 mag error
for $r^*=15-18$ mag) with the slope of the Schechter function 
$\alpha=-1.2$. Therefore, the values given in eqs. (\ref{eq:re-deV}) and
(\ref{eq:re-exp}) represent the average sizes for the $L^*$ galaxies. 

Our data indicate that the mean $h$ derived from late type disks 
is about the same value as the mean $r_e$ of early type galaxies. 
With these two parameters, our data are
consistent with the relations derived by Kent (1985).

%------------------------------------------------------------------

\section{Concentration index}

We define the (inverse) concentration index
by the ratio of the two Petrosian radii $C=r_{50}/r_{90}$ measured
in the $r'$ band.
Figure 10 shows $C$ against $T$ (see also Table 3). 
We have dropped 30 galaxies having $r_{50} < 2''$, 
since large smearing effects are suspected for $C$ for such small 
galaxies unless
significant corrections are made for the seeing (average 
FWHM is $\sim 1.''5$).
The two filled squares with error bars near the top of the panel 
show the typical errors in $T({\rm ours})$ measurements 
at $T=0$ and 3.

\setcounter{figure}{9}
\begin{figure}[t]
\plotone{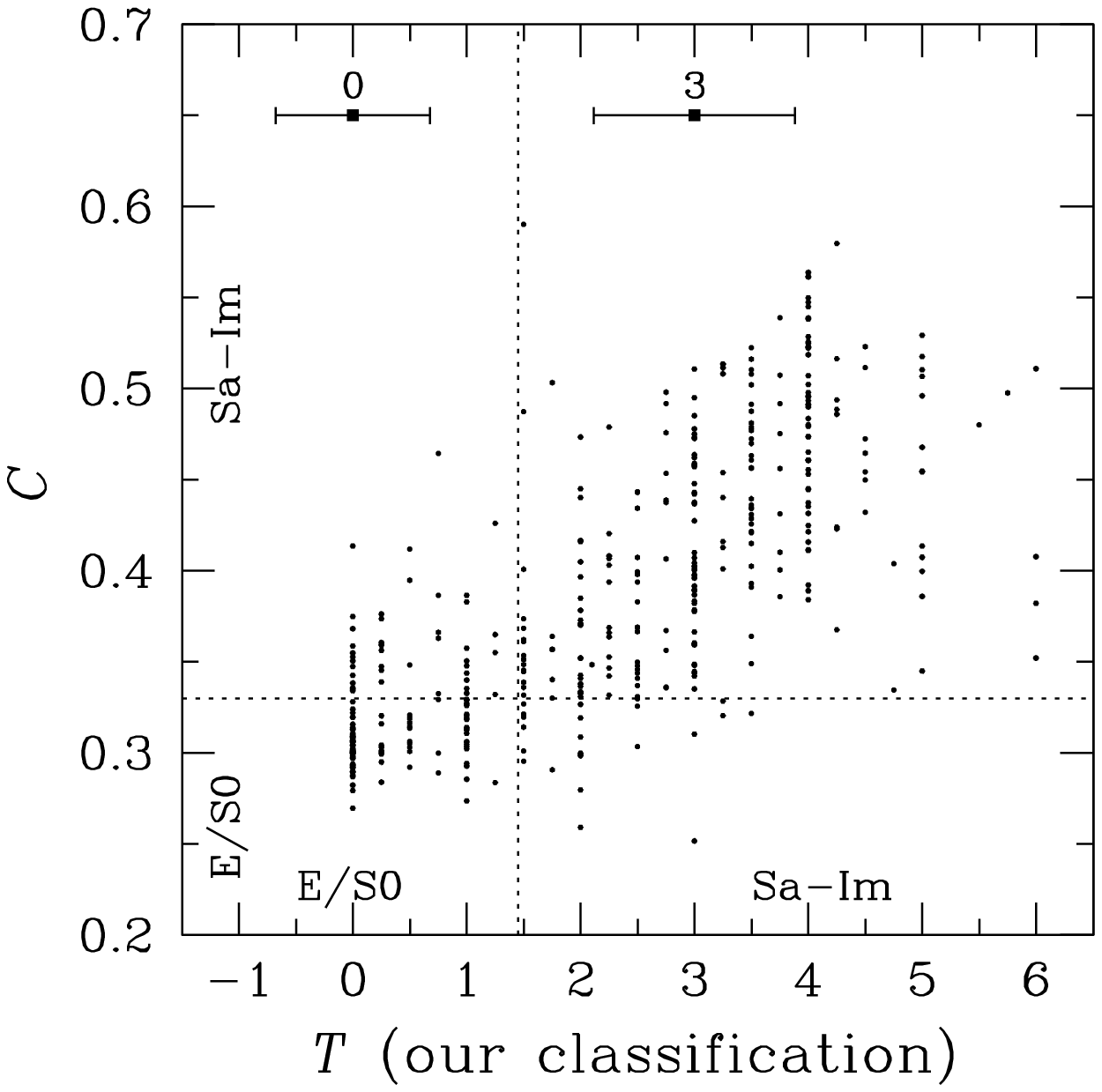}
\caption
{Concentration index $C$ plotted against eye morphology $T$
for 426 galaxies having $r_{50} \ge 2''$.
The vertical line indicates the boundary which divides
E/S0 galaxies and spiral/Im galaxies on the basis of
eye morphology.
The horizontal line is for the boundary accepted in the
text ($C_1 = 0.33$) dividing
low-$C$ (`early type') and high-$C$ (`late type') galaxies.
The two filled squares with error bars near the top of the panel 
show the errors in $T({\rm ours})$ measurements at $T=0$ and 3 
estimated from the difference between our classification and that 
of RC3.
\label{fig10}}
\end{figure}

A tight correlation is seen between $C$ and $T$;
$C$ increases with $T$, which is understood by
the spheroid or bulge dominance of early type galaxies and 
an increasing disk dominance 
of later-type galaxies.
We have studied similar plots for galaxies with magnitudes divided
into three bins, $r^*<16$, $16<r^*<17$, and $17<r^*<18$. No systematic
effects are observed among the samples with different apparent 
brightnesses (hence different apparent sizes) of galaxies, in so far 
as we limit ourselves to galaxies with  
$r_{50} \ge 2''$.  The galaxies with 
$r_{50} < 2''$ show a weaker correlation between $C$ and $T$
due to smearing by seeing but also possibly due to
errors of eye classification.
\footnote{
We have examined whether highly inclined galaxies would disturb  
the correlations between $C$ and $T$.
We have found that the correlation does not change appreciably
againt the sample with specified minor-to-major axis ratio.
Our concentration index defined using the Petrosian radii is a robust 
indicator against inclination of galaxies.
}

The scatter in $C$ for a given $T$ is larger than 
expected from the errors in our $T$ measurements.
This reflects the fact that galaxies with a given morphology 
have a fairly wide range of the bulge-to-disk ratio 
(e.g., Kent 1985; Simien \& de Vaucouleurs 1986; 
Kodaira, Watanabe, \& Okamura 1986)
and the bulge-to-disk ratio is directly related 
to the concentration index.

We examined the correlations of visual morphology with 
a number of parameters measured by {\it Photo} 
(likelihoods of a fitting to the de Vaucouleurs or 
exponential profiles, texture parameters defined by {\it Photo})  
in addition to the three parameters studied in this paper.
The correlation of the likelihood of profile fitting and of the texture
parameter with our visual morphology is not strong
(we used the version of {\it Photo} version\ $5\_0\_3$ of late 1999).
The concentration index shows the strongest correlation 
with visual morphology, 
and a combination of the concentration index with other parameters,
such as surface brightness, color, asymmetry parameter, 
does not appreciably enhance the correlation.

The concentration index is perhaps the best parameter
to be used to classify morphology of galaxies. This is
basically the same conclusion as reached by Doi et al. (1993),
and further corroborated by Abraham et al. (1994).
These authors used a combination of a concentration index and mean 
surface brightness, while the concentration index is
sufficient for our case. This simplification arises merely from the
different definition of the concentration indices.
In Doi et al. and Abrahams et al. 
the concentration index is defined using isophotal photometry. 
The calculated index then depends on both 
apparent brightness and distance of galaxies.
To correct for these redundant effects the second parameter
is needed, and those authors took average surface brightness,
which also depends on apparent brightness and the distance.
In our case the Petrosian
quantities are physically well defined, independent of the
apparent brightness or the distance.
Therefore the introduction of a second parameter is unnecessary.

The correlation in Figure 10 suggests that the $C$ parameter can
be used for morphological classification allowing for the uncertainty
of roughly $\Delta T\sim 1.5$ (see Table 3 above).
The most important problem in the classification using the $C$
parameter is that the $C$ parameter for Sa galaxies 
can be as small as that for E galaxies, 
making it difficult to construct a purely early type 
galaxy sample free from a contamination of Sa galaxies.
As noted above this problem is not ameliorated with the use of 
any second parameters we studied. 
On the other hand, the 
distinction between S0 and Sa is not too difficult, allowing for
some intermediate cases, upon visual inspections due to the
presence of spiral arms and/or HII regions in Sa galaxies,
which are not properly picked up with the parameters given by
{\it Photo}.  

A practically important application is to distinguish E/S0 galaxies
from spiral galaxies. Here we consider this problem  
in a more quantitative way, and study the
completeness and contamination of the morphologically classified sample
with the use of the $C$ parameter. 
The upper panel of Figure 11 shows the completeness as a function
of $C$: the solid curve represents the completeness of early type
galaxies when sample is selected with $C<C_1$ for a given $C_1$, 
while the dashed curve represents 
the completeness of the late type galaxy sample with $C \ge C_1$. 
The completeness of the two samples balances with $C_1=0.35$ at 80\%. 
On the other hand, the lower panel of Figure 11 indicates
the contamination from the opposite type. 
Namely, the solid curve is the contamination from late type galaxies 
to the early type galaxy sample with $C<C_1$. The dashed 
curve is the contamination by early type galaxies to the late type
sample with $C>C_1$. 
The two curves become even at $C_1=0.32$.

\setcounter{figure}{10}
\begin{figure}[t]
\plotone{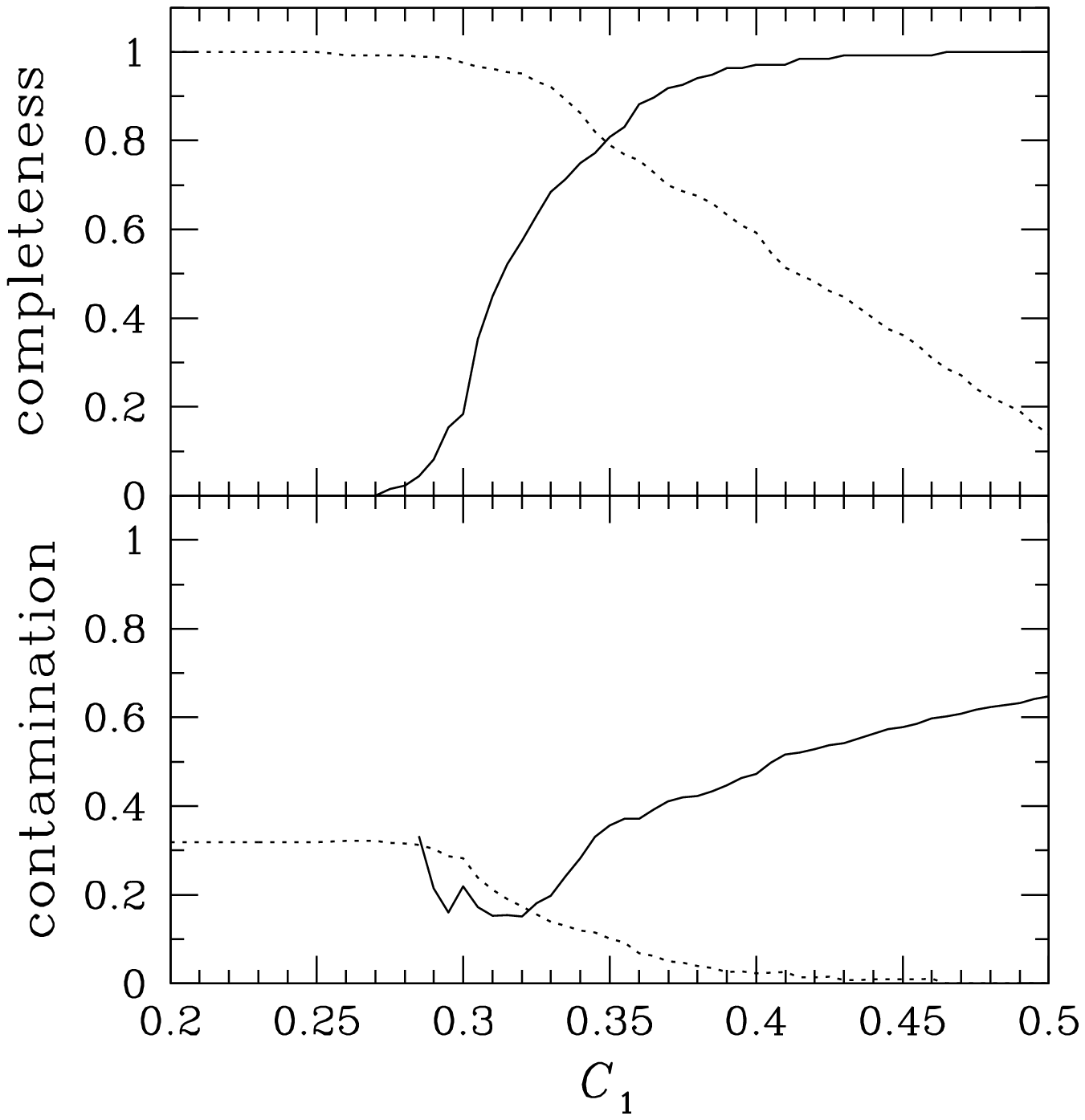}
\caption
{ Completeness and contamination of early and late type
galaxies as a function of $C_1$.
The solid and dotted lines are for early and late type
galaxies, respectively.
\label{fig11}}
\end{figure}

The inspection of two curves in the lower panel of Figure 11
shows that we can make a reasonably pure
late type galaxy sample with the choice of large $C_1$, say, 0.37
at a loss of completeness (70\% completeness; contamination is less 
than 5\%). The opposite is not true: one cannot decrease the
contamination of an early type sample by late type galaxies 
below 15\%, whatever value of $C_1$ is chosen,
as a result of a contamination from 
Sa galaxies\footnote{We suspect that this contamination is not due
to Seyfert galaxies, 
since the color of the contaminants are not bluer than others.}.

Our recommended choice of $C_1$ is 0.33 if we want dichotomous
classification of galaxies into early and late types. For this
specific choice the completeness and contamination are given
in Table 4.  
93 out of the 136 
early type (E/S0) galaxies are classified as `early' types, 
i.e., a completeness of $68 \%$.
On the other hand, 24 `early'-type galaxies are not real
E/S0 galaxies, which means a contamination of $21 \%$ (24/117).
Similarly, the fraction of late type (Sa-Im) galaxies classified 
as `late' types is $92 \%$ (266/290) and the contamination in 
the `late' galaxy sample from early type galaxies is $14 \%$ (43/309).

In summary, our analysis demonstrates that the parameter 
$C=r_{50}/r_{90}$ is useful for automated classification 
of early and late type galaxies, if one is satisfied with 
a completeness of $\simeq 70-90\%$, allowing for a contamination 
of $\simeq 15-20\%$. It appears difficult to improve the performance
with the use of simple photometric parameters. Further improvement
requires the introduction of more complicated parameters such as those
characterizing the presence of spiral arms and HII regions.
An improvement of late type galaxy classification may be 
achieved with the introduction
of the asymmetry parameter (e.g., Abraham et al. 1996). 

%------------------------------------------------------------------

\section{Conclusions}

We have produced a morphologically classified sample of 456 bright 
galaxies from imaging data which were taken during 
the Sloan Digital Sky Survey commissioning phase.
We have classified galaxies into seven Hubble types 
(E, S0, Sa, Sb, Sc, Sdm, and Im) by visual 
inspections carried out by four individuals.
We have measured the distributions of colors, scale lengths, 
and concentration indices in the SDSS photometric system  
as functions of morphology. Our prime purposes are (i)
to derive the fundamental statistical properties of
galaxies in the SDSS passbands, (ii) to examine the consistency
with existing galaxy data other than the SDSS, and (iii) to find
the parameters that can be used for an efficient automated 
morphological classifier.  
All parameters we have studied in this paper, 
except for visually determined morphology, 
are among the standard outputs 
from the SDSS image analysis software ({\it Photo}), 
and thus routinely registered in the SDSS photometric catalog.

We have found that the colors of our galaxies in the SDSS system 
match with those obtained from 
template SED data and from $UBVR_CI_C$ broad band photometry of 
nearby galaxies to within $\approx$0.05 mag.
The exception is
$i'-z'$ color for which the discrepancy is as large as 0.1-0.2 mag.
We could not find the origin of this discrepancy yet.

We found that the half light radius falls on the
line $\log r_{50} = -0.2m + {\rm const}$, and the constant we found
indicates $r_e\simeq 2.6$ kpc for de Vaucouleurs profiles, and
$h\simeq 2.6$ kpc for exponential disks of late type spiral galaxies
with the aid of the surface brightness radius relation derived by
Kent (1985).
We found additionally that the half light radius of galaxies 
depends slightly on
the color band and the dependence is consistent with the expected
distribution of star forming regions for late type galaxies 
and with the known color gradient in early type galaxies.  

Finally, we have shown that the concentration index
shows a strong correlation with the morphological types.
We explored the possibility of using this parameter 
for automated morphological classification of galaxies, 
in order to classify a huge number of galaxies 
into early, intermediate, and late types 
with parameters that are routinely produced in the photometric
data processing of the SDSS. 
For an optimum ratio of $[r_{50}/r_{90}]_1 \equiv C_1=0.33$, 
we find, however, a $15-20\%$ contamination 
from opposite types of galaxies. 
The contamination is particularly important for the
early type galaxy sample, and we must consider some higher
order photometric parameters to enhance the performance.

\vspace{10pt}

\noindent
{\it Note Added}

{\it Photo} has been continuously updated over the years.
After we submitted our paper, Strateva et al. (submitted to AJ)
found 
on the basis of a different sample (287 galaxies at $g^*<16$) 
that the likelihood of profile fitting based 
on a later version of {\it Photo} ($5\_2$) correlates 
with eye morphology better than we found.
However, they also found that 
the profile fitting does not work well in comparison with 
concentration index 
for bright galaxies such as those we study in our paper.

\vspace{10pt}

\noindent

We thank the referee, Roberto Abraham, for his valuable comments.

The Sloan Digital Sky Survey (SDSS) is a joint project of The 
University of Chicago, Fermilab, the Institute for Advanced Study, 
the Japan Participation Group, The Johns Hopkins University, 
the Max-Planck-Institute for Astronomy (MPIA), 
the Max-Planck-Institute for Astrophysics (MPA), 
New Mexico State University, Princeton University, the United States 
Naval Observatory, and the University of Washington. Apache Point 
Observatory, site of the SDSS telescopes, is operated 
by the Astrophysical Research Consortium (ARC). 
Funding for the project has been provided by the Alfred P. Sloan 
Foundation, the SDSS member institutions, the National Aeronautics 
and Space Administration, the National Science Foundation, 
the U.S. Department of Energy, the Japanese Monbukagakusho, 
and the Max Planck Society. 
The SDSS Web site is http://www.sdss.org/.

%------------------------------------------------------------------

%------------------------------------------------------------------

\clearpage

\begin{table}
\caption{Mean colors of SDSS galaxies as a function of $T$.}
\begin{tabular}{ccllll}
\hline
     $T$          & Hubble type
  & $u^*-g^*$    & $g^*-r^*$    & $r^*-i^*$    & $i^*-z^*$ \\
\hline
$T < 0.5$         & E
  & 1.79 (0.26)& 0.83 (0.14)& 0.41 (0.05)& 0.27 (0.06) \\
%& & 1.77       & 0.83       & 0.41       & 0.28        \\
& & 86         & 87         & 87         & 87          \\
$0.5\le T < 1.5$  & S0
  & 1.66 (0.30)& 0.75 (0.14)& 0.38 (0.05)& 0.26 (0.06) \\
%& & 1.71       & 0.78       & 0.39       & 0.28        \\
& & 67         & 67         & 67         & 67          \\
$1.5\le T < 2.5$  & Sa
  & 1.49 (0.32)& 0.68 (0.15)& 0.38 (0.07)& 0.25 (0.08) \\
%& & 1.51       & 0.72       & 0.38       & 0.25        \\
& & 81         & 81         & 81         & 81          \\
$2.5\le T < 3.5$  & Sb
  & 1.40 (0.28)& 0.62 (0.13)& 0.35 (0.08)& 0.20 (0.09) \\
%& & 1.36       & 0.63       & 0.35       & 0.21        \\
& & 95         & 94         & 93         & 94          \\
$3.5\le T < 4.5$  & Sc
  & 1.28 (0.33)& 0.46 (0.12)& 0.27 (0.09)& 0.07 (0.16) \\
%& & 1.26       & 0.46       & 0.27       & 0.11        \\
& & 94         & 98         & 97         & 91          \\
$4.5\le T < 5.5$  & Sdm
  & 1.16 (0.38)& 0.42 (0.24)& 0.20 (0.12)& 0.04 (0.20) \\
%& & 1.08       & 0.35       & 0.19       & 0.08        \\
& & 22         & 22         & 21         & 18          \\
$5.5\le T$     & Im
  & 1.14 (0.33)& 0.61 (0.23)& 0.23 (0.08)& 0.05 (0.17) \\
%& & 1.00       & 0.59       & 0.19       & 0.14        \\
& & 6         &  6         &  5         &  5          \\
\hline
\end{tabular}
\label{tab1}

\tablecomments{
The numbers in the parentheses are dispersions.
The numbers in the second rows for each morphology listing
are the number of galaxies used to calculate the mean.
Only galaxies satisfying the following criteria are used to
calculate the mean:
$0    \le u^*-g^* \le 2.5$,
$0    \le g^*-r^* \le 1.5$,
$0    \le r^*-i^* \le 1$,
$-0.5 \le i^*-z^* \le 0.5$.
}

\end{table}

\begin{table}
\caption{
Fit coefficients $a_{ij}$ in the relation between apparent 
half light radius and apparent brightness. 
}
\begin{tabular}{ccccccc}
\hline
     & $\lambda_i$ & $\lambda_j$ &     All & $T<1.5$ & $1.5\le T<4$ 
& $T \ge 4$ \\
\hline
1 & $u'$ & $r'$ &  0.51 (0.19) &  0.37  (0.15) &   0.56 (0.16) & 0.63 
(0.20)     \\
2 & $g'$ & $r'$ &  0.51 (0.18) &  0.36  (0.12) &   0.55 (0.14) & 0.68 
(0.17)     \\
3 & $r'$ & $r'$ &  0.49 (0.18) &  0.36  (0.12) &   0.52 (0.14) & 0.66 
(0.17)     \\
4 & $i'$ & $r'$ &  0.48 (0.17) &  0.35  (0.12) &   0.51 (0.14) & 0.64 
(0.16)     \\
5 & $z'$ & $r'$ &  0.44 (0.16) &  0.33  (0.12) &   0.47 (0.14) & 0.57
(0.16)      \\
6 & $u'$ & $u'$ &  0.94 (0.16) &  0.87 (0.15)  &   0.96 (0.14) & 0.99 
(0.17)  \\
7 & $g'$  & $g'$&  0.64 (0.16) &  0.52  (0.13) &   0.67 (0.13) & 0.77 
(0.16)  \\
8 & $r'$ & $r'$ &  0.49 (0.18) &  0.36  (0.12) &   0.52 (0.14) & 0.66 
(0.17)   \\
9 & $i'$ & $i'$ &  0.41 (0.18) &  0.27  (0.12) &   0.44 (0.14) & 0.59 
(0.17)  \\
10 & $z'$ & $z'$&  0.34 (0.19) &  0.20  (0.12) &   0.36 (0.16) & 0.53 
(0.19)     \\
\hline
\end{tabular}
\label{tab2}

\tablecomments{
See eq. (5) for the definition. 
The numbers in parentheses are dispersion around the fit.
}

\end{table}

\clearpage

\begin{table}
\caption{
Concentration index $C=r_{50}/r_{90}$.
}
\begin{tabular}{crrcc}
\hline
$T({\rm our})$      & $N$ & & \multicolumn{2}{c}{$C$} \\
                    &     & &  mean   &    rms    \\
\hline
$T < 0.5         $  & 80  & & 0.317   &  0.027   \\
$0.5 \le T < 1.5 $  & 56  & & 0.331   &  0.037   \\
$1.5 \le T < 2.5 $  & 73  & & 0.362   &  0.054   \\
$2.5 \le T < 3.5 $  & 92  & & 0.404   &  0.057   \\
$3.5 \le T < 4.5 $  & 97  & & 0.467   &  0.052   \\
$4.5 \le T < 5.5 $  & 22  & & 0.452   &  0.055   \\
$5.5 \le T       $  &  6  & & 0.439   &  0.061   \\
\hline
\end{tabular}
\label{tab3}
\end{table}

\begin{table}
\caption{
Correlation between eye morphology and that classified with
the concentration index with $C_1=0.33$.
}
\begin{tabular}{crrrc}
\hline
             & $T < 1.5$ & $T\ge1.5$ &  sum   &  contamination \\
\hline
`early type' galaxies   &   93  &   24  &  117  &  21 \%  \\
`late type' galaxies    &   43  &  266  &  309  &  14 \%  \\
sum                     &  136  &  290  &  426  &         \\
\hline
completeness            &  68\% &  92\% &  ---  &   ---   \\
\hline
\end{tabular}
\label{tab4}
\end{table}


\begin{thebibliography}{}

\bibitem{}
Abraham, R. G., Tanvir, N. R., Santiago, B. X., Ellis, R. S., 
Glazebrook, K., \& van den Bergh, S.
1996, \mnras, 279, L47

\bibitem{}
Abraham, R. G., Valdes, F., Yee, H. K. C., \& 
van den Bergh, S.
1994, \apj, 432, 75

\bibitem{}
Bruzual, A. G. \& Charlot, S. 
1993, \apj, 405, 538

\bibitem{}
Buta, R. \& Williams, K. L.
1995, \aj, 109, 543 (BW)

\bibitem{}
Coleman, G. D., Wu, C.-C., \& Weedman, D. W.
1980, \apjs, 43, 393 (CWW)

\bibitem{}
Doi, M., Fukugita, M., \& Okamura, S.
1993, \mnras, 264, 832

\bibitem{}
Frei, Z. \& Gunn, J. E.
1994, \aj, 108, 1476

\bibitem{}
Fukugita, M., Hogan, C. J., \& Peebles, P. J. E.
1998, \apj, 503, 518

\bibitem{}
Fukugita, M., Ichikawa, T., Gunn, J. E., Doi, M., Shimasaku, K., 
\& Schneider, D. P.
1996, \aj, 111, 1748

\bibitem{}
Fukugita, M., Shimasaku, K., \& Ichikawa, T.
1995, \pasp, 107, 945 (FSI)

\bibitem{}
Gunn, J. E. \& Stryker, L. L.
1983, \apjs, 52, 121

\bibitem{}
Gunn, J. E., et al.
1998, \aj, 116, 3040

\bibitem{}
Kennicutt, R. C. 1992, \apjs, 79, 255

\bibitem{}
Kent, S. M.
1985, \apjs, 59, 115

\bibitem{}
Kodaira, K., Watanabe, M., \& Okamura, S.
1986, \apjs, 62, 703

\bibitem{}
Kodama, T. \& Arimoto, N.
1997, \aap, 320, 41

%neural network
\bibitem{}
Lahav, O., Naim, A., Sodr\'e, L., Jr., \& Storrie-Lombardi, M. C.
1996, MN, 283, 207

\bibitem{}
Loveday, J.
1996, \mnras, 278, 1025

\bibitem{}
Morgan, W. W. 
1958, \pasp, 70, 364

\bibitem{}
% Naim, A., Lahav, O., Buta, R. J., Corwin, H. G., Jr., 
% de Vaucouleurs, G., Dressler, A., Huchra, J. P., van den Bergh, S., 
% Raychaudhury, S., Sodre, L., Jr., Storrie-Lombardi, M. C.
Naim, A. et al.
1995, \mnras, 274, 1107

\bibitem{}
Sandage, A. 
1961, The Hubble Atlas of Galaxies.
Carnegie Institution of Washington, Washington DC

\bibitem{}
Schlegel, D. J., Finkbeiner, D. P. \& Davis, M.
1998, \apj, 500, 525

\bibitem{}
Simien, F. \& de Vaucouleurs, G.
1986, \apj, 302, 564

\bibitem{}
Smith, J. A. 2001, AAS meeting: 197.1311S


\bibitem{}
de Vaucouleurs, G.
1948, Ann. d'Ap., 11, 247

\bibitem{}
de Vaucouleurs, G., de Vaucouleurs, A., Corwin, H., 
Buta, R., Paturel, G., \& Fouqu\'e, P.
1991, 
Third Reference Catalogue og Bright Galaxies 
(Springer, New York) (RC3)

\bibitem{}
Yasuda, N., et al.
2000, submitted to AJ

\bibitem{}
York, D. G., et al.
2000, \aj, 120, 1579

\end{thebibliography}
\end{document}